# Noise effects and perfect controlled remote state preparation

Yuan-hua Li[1], Zisheng Wang[2], Hua-qing Zhou[1], Xian-feng Chen[1]

[1] The State Key Laboratory on Fiber Optic Local Area Communication Networks and Advanced Optical Communication Systems, Department of Physics and Astronomy, Shanghai Jiao Tong University, Shanghai 200240, China

[2] Department of physics, Jiangxi Normal University, Nanchang 330022, China

**Abstract:** We investigate all possible noise environments for controlled remote state preparation. We find that the optimal efficiency is not only dependent on the isolate decay rate but also dependent on the coupling term of environments. Such a coupling provides a clue to control the noisy quantum system for realistic quantum communication. We show that, furthermore, the noise channel can be used as an important resource for quantum information tasks. Especially, an approach is proposed to realize a perfect controlled remote state preparation under the noisy effects environments by choosing the noisy environments and adjusting their relations in terms of noisy rates.

## 1. Introduction

Transmission of quantum states is a central task in quantum information science. If a sender (Alice) wants to transmit an unknown quantum state to a receiver (Bob), they can use quantum teleportation (QT) [1-3]. As shown in a remote state preparation protocol (RSP), however, the classical communication sending a known state is less cost than the teleportation [4-7]. Similarly to preparing Bob's qubit in a particular state determined by Alice, conditional on the outcome of a measurement on her qubit, RSP is a quantum communication protocol that relies on correlations between two entangled qubits. Unlike the teleportation, on the other hand, RSP does not require for the sender to perform full Bell-state analysis, currently an experimental challenge for optical implementations. Due to its interesting properties, RSP has been widely investigated both theoretically [8-16] and experimentally [17-25] in recent years.

As we know, one of the key factors in the perfect controlled RSP is to use an entangled pure quantum channel [26-28]. In any realistic RSP protocol, unfortunately, a quantum system interacts irreversibly with its reservoir [29-31]. Such an effect of noise or decoherence is to turn pure state into mixed one, and perfect controlled RSP is not possible since, instead of entangled pure states, noise or decoherence effects force us to deal with mixed states. And it has been shown that with classical communication and local operations perfect controlled RSP with mixed states is impossible in both theoretically [32] and experimentally [24] and therefore, one of the current major challenges in accomplishing perfect controlled RSP is to overcome the limitations imposed by noise.

Previously, environment-assisted quantum processes have been studied [33-37], and the results show that noise or decoherence can enhance the efficiency of quantum communication protocols. In this paper, we investigate how the efficiency is affected in the noisy environments in the controlled RSP protocol, where all possible types of noise or decoherence effects are considered in realistic quantum communication protocols. We discuss the several realistic scenarios, i.e., a part or all of the qubits are subjected to the same or different types of noise. Under several realistic situations, we show that the average fidelity can reach one, i.e., perfect controlled RSP can be achieved by adjusting the initial angle of the quantum channel and the controller's measurement basis vector, and controlling the noise rate or choosing the types of noisy environments. Thus it is possible to conquer the decrease of efficiency in the protocol due to the noise with another noise. In this work, we also consider all possible noise channels encountered in the laboratory as well as different scenarios in which one, two, or all three qubits are employed in the controlled RSP protocol interacting with the another noise.

## 2. Controlled remote state preparation protocol

Suppose that two participants, Alice and Bob, help for the remote receiver Charlie to prepare an arbitrary single-qubit state $|\psi\rangle = |\alpha||0\rangle + |\beta|e^{i\delta}|1\rangle$, where $|\alpha|^2 + |\beta|^2 = 1$ with the absolute values $|\alpha|$ and $|\beta|$ of the constant coefficients $\alpha$ and $\beta$ and the relative phase $\delta$. Here, Charlie prepares a three-qubit state $|\Phi^\theta\rangle_{123} = \cos\theta|000\rangle_{123} + \sin\theta|111\rangle_{123}$, where the qubit 1 belongs to Alice, qubit 2 belongs to Bob and qubit 3 belongs to Charlie, respectively. Thus $\theta = \pi/4$ is a three qubits GHZ state, i.e., a maximally entangled pure state, where the initial angle $0 \leq \theta \leq \pi/2$ of such a quantum entangled channel will be taken as a free parameter in order to fit the maximum efficiency of noisy controlled RSP protocol.

Next, it is necessary to define two orthogonal states in order to project the qubit onto Alice with her,

$$|A_1\rangle_1 = |\alpha||0\rangle + |\beta|e^{-i\delta}|1\rangle, \quad |A_2\rangle_1 = |\beta|e^{-i\delta}|0\rangle - |\alpha||1\rangle, \tag{1}$$

where the Bob's measurement basis vector is

$$|B_1^\varphi\rangle_2 = \cos\varphi|0\rangle + \sin\varphi|1\rangle, \quad |B_2^\varphi\rangle_2 = \sin\varphi|0\rangle - \cos\varphi|1\rangle, \tag{2}$$

under the region of $0 \leq \varphi \leq \pi/2$. When $\varphi = \pi/4$, the two single-qubit projective states are $|\pm\rangle = (|0\rangle \pm |1\rangle)/\sqrt{2}$. Under the noisy environment, the controllable free parameter $\varphi$ would be adjusted to optimize the efficiency of controlled RSP. In terms of Alice's and Bob's measurement basis vectors, the quantum channel can be rewritten as

$$\begin{aligned}|\Phi^\theta\rangle_{123} &= |A_1\rangle_1|B_1^\varphi\rangle_2 \left(|\alpha|\cos\theta\cos\varphi|0\rangle + |\beta|e^{i\delta}\sin\theta\sin\varphi|1\rangle\right)_3 \\ &+ |A_1\rangle_1|B_2^\varphi\rangle_2 \sigma_3^z \left(|\alpha|\cos\theta\sin\varphi|0\rangle + |\beta|e^{i\delta}\sin\theta\cos\varphi|1\rangle\right)_3 \\ &+ |A_2\rangle_1|B_1^\varphi\rangle_2 \sigma_3^z\sigma_3^x \left(|\alpha|\sin\theta\sin\varphi|0\rangle + |\beta|e^{i\delta}\cos\theta\cos\varphi|1\rangle\right)_3 \\ &+ |A_2\rangle_1|B_2^\varphi\rangle_2 \sigma_3^x \left(|\alpha|\sin\theta\cos\varphi|0\rangle + |\beta|e^{i\delta}\cos\theta\sin\varphi|1\rangle\right)_3. \end{aligned} \tag{3}$$

In order to realize the controlled RSP, Alice firstly performs a single-qubit measurement on the qubit 1 and publicly announces her measurement outcome. According to the Alice's measurement result, next, Bob should measure his qubit by choosing one of the measuring basis vectors. After the measurement, Bob informs Charlie about his measured result by using a classical channel. In terms of Alice's and Bob's measured results, Charlie can recover the desired state as shown in Eq. (3) by using a suitable unitary operation. Thus the successful probabilities $Q_j\{j=1,2,3,4\}$ can be obtained by

$$Q_1 = |\alpha|^2 \cos^2\theta\cos^2\varphi + |\beta|^2 \sin^2\theta\sin^2\varphi, \tag{4}$$

$$Q_2 = |\alpha|^2 \cos^2\theta\sin^2\varphi + |\beta|^2 \sin^2\theta\cos^2\varphi, \tag{5}$$

$$Q_3 = |\alpha|^2 \sin^2\theta\sin^2\varphi + |\beta|^2 \cos^2\theta\cos^2\varphi, \tag{6}$$

$$Q_4 = |\alpha|^2 \sin^2\theta\cos^2\varphi + |\beta|^2 \cos^2\theta\sin^2\varphi. \tag{7}$$

It is convenient to quantify the protocol efficiency in terms of the fidelity [38]. Since the benchmark state is an initially pure one, the fidelity $F_j \{j=1,2,3,4\}$ can be written as

$$F_1 = \frac{|\alpha|^4 \cos^2\theta \cos^2\varphi + |\beta|^4 \sin^2\theta \sin^2\varphi + 2|\alpha|^2 |\beta|^2 \cos\theta \cos\varphi \sin\theta \sin\varphi}{|\alpha|^2 \cos^2\theta \cos^2\varphi + |\beta|^2 \sin^2\theta \sin^2\varphi}, \quad (8)$$

$$F_2 = \frac{|\alpha|^4 \cos^2\theta \sin^2\varphi + |\beta|^4 \sin^2\theta \cos^2\varphi + 2|\alpha|^2 |\beta|^2 \cos\theta \cos\varphi \sin\theta \sin\varphi}{|\alpha|^2 \cos^2\theta \sin^2\varphi + |\beta|^2 \sin^2\theta \cos^2\varphi}, \quad (9)$$

$$F_3 = \frac{|\alpha|^4 \sin^2\theta \sin^2\varphi + |\beta|^4 \cos^2\theta \cos^2\varphi + 2|\alpha|^2 |\beta|^2 \cos\theta \cos\varphi \sin\theta \sin\varphi}{|\alpha|^2 \sin^2\theta \sin^2\varphi + |\beta|^2 \cos^2\theta \cos^2\varphi}, \quad (10)$$

$$F_4 = \frac{|\alpha|^4 \sin^2\theta \cos^2\varphi + |\beta|^4 \cos^2\theta \sin^2\varphi + 2|\alpha|^2 |\beta|^2 \cos\theta \cos\varphi \sin\theta \sin\varphi}{|\alpha|^2 \sin^2\theta \cos^2\varphi + |\beta|^2 \cos^2\theta \sin^2\varphi}. \quad (11)$$

Consider a controlled RSP protocol occurred in each state in terms of the different probabilities, we define the average fidelity as $\bar{F} = \sum_{j=1}^{4} Q_j F_j$, with $Q_j$ is a successful probability of the Charlie state, i.e., any qubit is equally probable to be picked as an original state in the controlled RSP protocol. Thus we find

$$\bar{F} = |\alpha|^4 + |\beta|^4 + 2|\alpha|^2 |\beta|^2 \sin(2\theta)\sin(2\varphi), \quad (12)$$

where $\bar{F}$ depends on the original state and the initial angle $\varphi$. The average value $\langle \bar{F} \rangle = \frac{1}{2\pi} \int_0^{2\pi} \int_0^1 \bar{F}(|\alpha|^2, \delta) d|\alpha|^2 d\delta$, and therefore the quantity

$$\langle \bar{F} \rangle = \frac{2}{3} + \frac{1}{3}\sin(2\theta)\sin(2\varphi) \quad (13)$$

is an efficiency of the controlled RSP protocol. Under the condition of $\theta = \varphi = \pi/4$, $\langle \bar{F} \rangle = 1$. The result implies that we recover the perfect controlled RSP protocol with $Q_1 = Q_2 = Q_3 = Q_4 = 1/4$ in terms of a maximally three-qubit pure entangled channel. Under the other situations, especially for the noises interacting with the quantum channel, $Q_j \{j=1,2,3,4\}$ depends on the desired state and an averaging over the other possible degrees of freedom in the controlled RSP protocol.

## 3. Noisy controlled RSP protocol

The interaction of a noisy environment with a qubit can be described by the quantum operators. In the operator-sum representation formalism, the trace-preserving Kraus operators

$E_k \{k=1,2,\cdots,n\}$ can represent such a noise and satisfy the complete condition,

$$\sum_{k=1}^{n} E_k^\dagger E_k = I, \qquad (14)$$

where $I$ is an identity operator acting on the qubit's Hilbert space. Under the noisy environment, the density matrix $\rho_j$ of the qubit $j$ becomes

$$\rho_j \to \varepsilon_j = \sum_{k=1}^{n} E_k^\dagger \rho_j E_k. \qquad (15)$$

In the following, we will discuss all possible types of noise from the realistic noisy environment as shown in Fig. 1, where one, two, or all three qubits of the quantum entangled pure channel in the controlled RSP protocol are affected by such a noise in a different way. And all possible types of noise are given in the Appendix.

Under case of that each qubit is independently subjected to the noisy environments, the initial density matrix will evolve in terms of the types of noise.

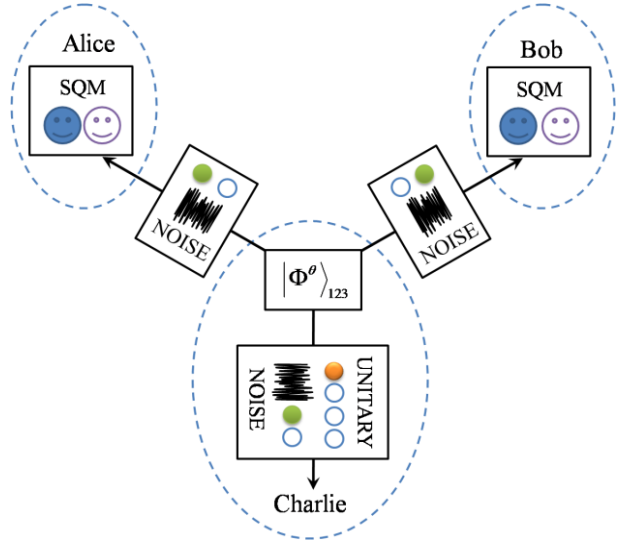

FIG. 1. (Color online) A schematic picture of the controlled RSP protocol. Three-qubit pure entangled channel $|\Phi^\theta\rangle_{123}$ refers to a source producing states of the form of Eq. (3) and SQM refers to a single-qubit measurement. The noise before the SQM makes the $|\Phi^\theta\rangle_{123}$ mixed while the noise in the last step of the protocol allows one to obtain the desired state.

From Eq. (15) with the three sources of noise, the density matrix including the noise effects is given by

$$\varepsilon = \sum_{i=1}^{n_1} \sum_{j=1}^{n_2} \sum_{k=1}^{n_3} E_{ijk}(p_1,p_2,p_3) |\Phi\rangle \langle\Phi| E_{ijk}^\dagger(p_1,p_2,p_3), \qquad (16)$$

where $E_{ijk}(p_1, p_2, p_3) = M_i(p_1) \otimes N_j(p_2) \otimes L_k(p_3)$, with $M_i(p_1) = M_i(p_1) \otimes I \otimes I$, $N_j(p_2) = I \otimes N_j(p_2) \otimes I$ and $L_k(p_3) = I \otimes I \otimes L_k(p_3)$. It is obvious that each Kraus operator is associated with one kind of the noise interacting on the desired qubit. Generally, the different noises can act during different times with different probabilities, which can be distinguished by $p_1$, $p_2$ and $p_3$. Inserting Eq. (16) into Eqs. (3)-(13), we could get the relevant physical quantities to analyze the perfect controlled RSP protocol under the noise environments. Under the limit of $p_1 = p_2 = p_3 = 0$, on the other hand, we can recover the noiseless case, i.e., the pure entangled state.

## 4. Discussions and Results

**A. Noise in Alice's qubit**

In order to make it clear which qubits are subjected to the noise, we introduce the notation $\langle \overline{F}_{X,\varnothing,Y} \rangle$ in terms of the optimal efficiency of the protocol, where the first subindex represents the qubit 1 interacting with the noise $X$, the second one denotes that Bob's qubit of the quantum channel without any noise, and the third subindex is Charlie's qubit interacting with the noise $Y$. Here $X$ and $Y$ can be any one of the four kinds of noise described in the appendix.

At the initial time, suppose that the qubits 2 and 3 of quantum channel are not affected from the noise, i.e., $p_2 = p_3 = 0$, and the qubit 1 lies in a noisy environment ($p_1 \neq 0$). The efficiency, for each type of noise as described in Sec. 3, can be obtained by

$$\langle \overline{F}_{BF,\varnothing,\varnothing} \rangle = \frac{2}{3}\left[1 - \frac{p_1}{2} + \frac{1}{2}\sin(2\theta)\sin(2\varphi)\right], \tag{17}$$

$$\langle \overline{F}_{AD,\varnothing,\varnothing} \rangle = \frac{2}{3}\left[1 - \frac{p_1}{2}\sin^2\theta + \frac{\sqrt{1-p_1}}{2}\sin(2\theta)\sin(2\varphi)\right], \tag{18}$$

$$\langle \overline{F}_{PhF,\varnothing,\varnothing} \rangle = \frac{2}{3}\left[1 + \frac{|1-2p_1|}{2}\sin(2\theta)\sin(2\varphi)\right], \tag{19}$$

$$\langle \overline{F}_{D,\varnothing,\varnothing} \rangle = \frac{2}{3}\left[1 - \frac{p_1}{4} + \frac{1-p_1}{2}\sin(2\theta)\sin(2\varphi)\right]. \tag{20}$$

In Eqs. (17) to (20), the subscripts on the left-hand side are the particular type of noise, i.e., $BF \rightarrow$ bit-flip, $AD \rightarrow$ amplitude-damping, $PhF \rightarrow$ phase-flip, and $D \rightarrow$ depolarizing.

From Eqs. (17)-(20), we see that the optimal efficiency is function of $p_1$, the initial angle $\theta$ of the quantum channel, and the initial angle $\varphi$ of the Bob's measurement basis vectors, where the maximum efficiencies are occurred at $\varphi = \pi/4$ due to the conditions $0 \leq \sin(2\theta)\sin(2\varphi) \leq 1$ and $1 - p_1 \geq 0$. In Fig. 2, we plot the numerical results of Eqs. (17)-(20), where $0 \leq p_1 \leq 1$, $0 \leq \theta \leq \pi/2$, and $\varphi = \pi/4$.

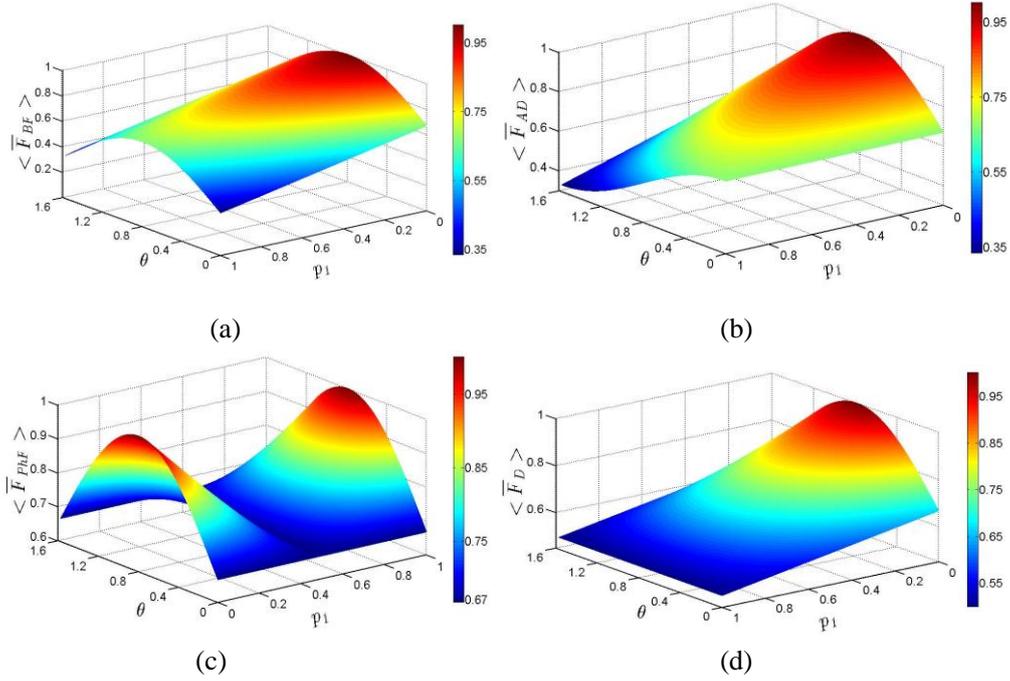

FIG. 2. (Color online) Efficiency of the controlled RSP protocol when only the desired qubit is affected by a noisy environment, with $p_1$ representing the probability for the noise to act on the qubit 1, and the initial angle $\theta$ of the quantum channel.

According to Fig. 2, the maximum efficiencies $\langle \overline{F}_{BF,\varnothing,\varnothing} \rangle = \langle \overline{F}_{AD,\varnothing,\varnothing} \rangle = \langle \overline{F}_{D,\varnothing,\varnothing} \rangle = 1$ occur at $p_1 = 0$ and $\theta = \pi/4$, i.e., the qubit 1 is not affected from the noise. It is surprised that the efficiency $\langle \overline{F}_{PhF,\varnothing,\varnothing} \rangle$ is separated into two the two regions. In the range $0 \leq p_1 \leq 1/2$, the average fidelity decreases and increases in the range of $1/2 \leq p_1 \leq 1$. Thus the maximum efficiency $\langle \overline{F}_{PhF,\varnothing,\varnothing} \rangle = 1$ occur at $p_1 = 0$ or $p_1 = 1$ for $\theta = \pi/4$. In the two cases of $p_1 = 0$ and $p_1 = 1$, therefore, the phase flip does not affect on the physical system. The results give out an approach to control the RSP protocol under the environment of phase flip. For high values of $p_1$, the phase-flip noise give the greatest efficiency in the case of $\theta = \varphi = \pi/4$.

Next, let us investigate a more real situation. We firstly study that the qubit 1 is always subjected to the bit-flip noise and Charlie's qubit lies in one of the four different types of noisy environments given in Sec. 3. The optimal efficiencies are given by

$$\langle \overline{F}_{BF,\varnothing,BF} \rangle = \frac{2}{3} - \frac{1}{3}(p_1 + p_3 - 2p_1 p_3) + \frac{1}{3}\sin(2\theta)\sin(2\varphi), \tag{21}$$

$$\langle \overline{F}_{BF,\varnothing,AD} \rangle = \frac{2}{3} - \frac{p_1}{3} - \frac{p_3}{3}(1-2p_1)\sin^2\theta + \frac{\sqrt{1-p_3}}{3}\sin(2\theta)\sin(2\varphi), \tag{22}$$

$$\langle \overline{F}_{BF,\varnothing,PhF} \rangle = \frac{2}{3} - \frac{1}{3}p_1 + \frac{|1-2p_3|}{3}\sin(2\theta)\sin(2\varphi), \tag{23}$$

$$\langle \overline{F}_{BF,\varnothing,D} \rangle = \frac{2}{3} - \frac{p_3}{6} - \frac{p_1}{3} + \frac{p_1 p_3}{3} + \frac{1}{3}(1-p_3)\sin(2\theta)\sin(2\varphi). \tag{24}$$

From Eqs. (21)-(24), we see the efficiencies depend on noisy rates $p_1$ and $p_3$ for the optimal efficiencies with $\theta = \varphi = \pi/4$, where the coupling terms between the noisy rates $p_1$ and $p_3$ in Eqs. (21)-(24). Such coupling terms imply the entanglement of noisy environments. Therefore, there exists a possible approach to obtain the maximum efficiencies by adjusting the values of the parameters $p_1$ and $p_3$ in the efficiencies $\langle \overline{F}_{BF,\varnothing,BF} \rangle$, $\langle \overline{F}_{BF,\varnothing,AD} \rangle$, $\langle \overline{F}_{BF,\varnothing,PhF} \rangle$ and $\langle \overline{F}_{BF,\varnothing,D} \rangle$. In Fig. 3, thus, we firstly plot the efficiencies for different values of the noisy rate $p_1$ and $p_3$ in terms of Eqs. (21)-(24).

For $p_1 < 0.5$, we find that $\langle \overline{F}_{BF,\varnothing,BF} \rangle$, $\langle \overline{F}_{BF,\varnothing,PhF} \rangle$ and $\langle \overline{F}_{BF,\varnothing,D} \rangle$ reduce with increasing of the noisy rate $p_3$. The results imply that the average fidelities decrease with increasing of noisy rate. Differently from $\langle \overline{F}_{BF,\varnothing,AD} \rangle$ and $\langle \overline{F}_{BF,\varnothing,D} \rangle$, $\langle \overline{F}_{BF,\varnothing,PhF} \rangle$ is divided into two regions, where $\langle \overline{F}_{BF,\varnothing,PhF} \rangle$ raises for $p_3 > 0.5$ and reduces for $p_3 < 0.5$.

Under the situations of $p_1 > 0.5$, the average fidelity $\langle \overline{F}_{BF,\varnothing,BF} \rangle$ becomes bigger with increasing of noisy rate $p_3$. The results show that more noise means higher efficiency in the cases. By putting Charlie's qubit in a noisy environment described by the bit-flip map, thus, we can raise the efficiency of the protocol beyond the classical limit with the value 2/3 for $p_1 > 0.5$.

In addition, we find that $\langle \overline{F}_{BF,\varnothing,BF} \rangle = 1$ for $p_1 = p_3 = 0$ or $p_1 = p_3 = 1$ with $\theta = \varphi = \pi/4$, $\langle \overline{F}_{BF,\varnothing,PhF} \rangle = 1$ for $p_3 = 0$ or $p_3 = 1$ with $\theta = \varphi = \pi/4$ and $p_1 = 0$, and $\langle \overline{F}_{BF,\varnothing,AD} \rangle = \langle \overline{F}_{BF,\varnothing,D} \rangle = 1$ for $p_1 = p_3 = 0$ with $\theta = \varphi = \pi/4$ as shown in Fig. 3.

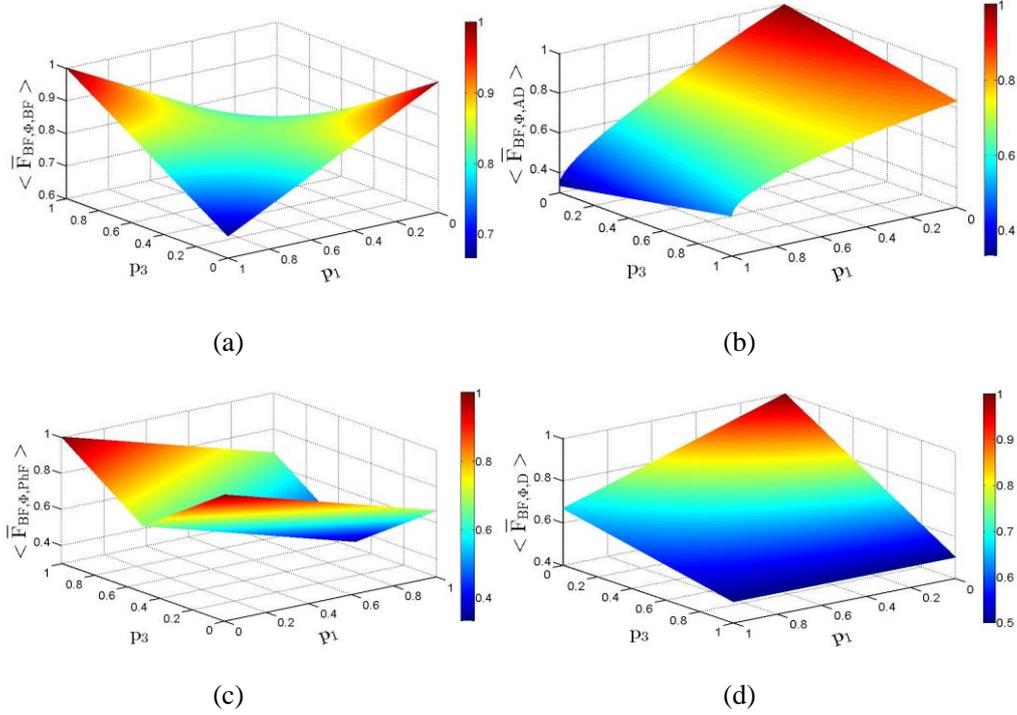

(a)  (b)

(c)  (d)

FIG. 3. (Color online) Efficiency of the controlled RSP protocol when both Alice's qubit ($p_1$) and Charlie's qubit ($p_3$) are affected by a noisy environment. Here the qubit 1 is always subjected to the bit-flip (BF) noise while Charlie's qubit may suffer from several types of noise.

In order to find a perfect controlled RSP under the noisy environment by controlling the noisy rates, we set $\langle \overline{F}_{BF,\varnothing,AD} \rangle = 1$ in Eq. (22). Thus we have

$$p_3 = \frac{1}{2\sin^2\theta(1-2p_1)^2}\left\{2\sin^2\theta p_1 - 2\sin^2\theta - \sin^2(2\theta) + 4\sin^2\theta p_1^2 \right. \qquad (25)$$
$$\left. + \sin(2\theta)\sqrt{4\sin^2\theta + 4\sin^4\theta + \sin^2(2\theta) - 4\sin^2\theta p_1 - 16\sin^2(2\theta)p_1 - 8\sin^2\theta p_1^2 + 16\sin^4\theta p_1^2}\right\}.$$

For $0 \leq p_1 \leq 1$, $0 \leq \theta \leq \pi/2$ in Eq. (25), if $p_3$ exist in the range $0 \leq p_3 \leq 1$, we can recover the perfect controlled RSP protocol, otherwise, we cannot realize the perfect controlled RSP. We plot the numerical results of Eq. (25), where $0 \leq p_1 \leq 1$ and $0 \leq \theta \leq \pi/2$. Fortunately, one can obtain $\langle \overline{F}_{BF,\varnothing,AD} \rangle = 1$ for $0.5 < p_1 \leq 0.6$, $0 \leq p_3 < 0.3$, and $0 < \theta \leq 1$ [rad] as shown in Fig. 4.

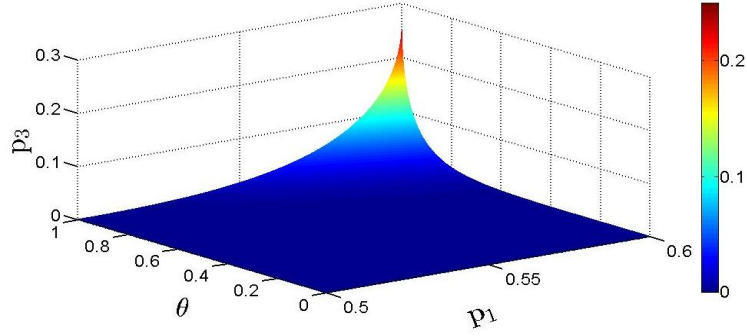

FIG. 4. (Color online) The relations of realizing the perfect controlled RSP between the noisy rates and measured angle, where the qubit 1 is subjected to the bit-flip (BF) noise while Charlie's qubit may suffer from amplitude-damping (AD) noise: $0.5 < p_1 \leq 0.6$, $0 \leq p_3 < 0.3$, $0 < \theta \leq 1$ [rad], and $\langle \overline{F}_{BF,\varnothing,AD} \rangle = 1$

The results show that one can implement the perfect controlled RSP by using a non-maximally three-qubit pure entangled state under the noise environment in terms of Eq. (22). For the other environments, unfortunately, we can not find a way to implement the perfect controlled RSP in terms of Eq. (21) and Eqs.(23)-(24).

Let us consider another case of qubit 1 interacting with the amplitude-damping noise, while Charlie's qubit can suffer any one of the four kinds of noise as shown in the appendix. The optimal efficiencies are now given by

$$\langle \overline{F}_{AD,\varnothing,BF} \rangle = \frac{2}{3} - \frac{p_3}{3} - \frac{p_1}{3} - \frac{p_1 p_3}{6} + \frac{p_1}{6}(1+p_3)\sin^2\theta + \frac{\sqrt{1-p_1}}{3}\sin(2\theta)\sin(2\varphi), \qquad (26)$$

$$\langle \overline{F}_{AD,\varnothing,AD} \rangle = \frac{2}{3} - \frac{1}{3}(p_1 + p_3 - 2p_1 p_3)\sin^2\theta + \frac{\sqrt{(1-p_1)(1-p_3)}}{3}\sin(2\theta)\sin(2\varphi), \qquad (27)$$

$$\langle \overline{F}_{AD,\varnothing,PhF} \rangle = \frac{2}{3} - \frac{p_1}{3} + \frac{p_1}{3}\sin^2\theta + \frac{|1-2p_3|\sqrt{1-p_1}}{3}\sin(2\theta)\sin(2\varphi), \qquad (28)$$

$$\langle \overline{F}_{AD,\varnothing,D} \rangle = \frac{2}{3} - \frac{p_3}{6} - \frac{p_1}{3} + \frac{p_1 p_3}{12} + \frac{p_1}{6}(2+p_3)\sin^2\theta + \frac{(1-p_3)\sqrt{1-p_1}}{3}\sin(2\theta)\sin(2\varphi). \qquad (29)$$

From Eqs. (26)-(29), we see that the efficiencies depend on noisy rate $p_1$, $p_3$, $\theta$ and $\varphi$, where the optimal efficiencies are occurred at $\theta = \varphi = \pi/4$. In Fig. 5, we plot the optimal efficiencies for different values of the noisy rate $p_1$ and $p_3$ in terms of Eqs. (26)-(29), where $0 \leq p_1 \leq 1$, $0 \leq p_3 \leq 1$, and $\theta = \varphi = \pi/4$.

In Fig. 5, we find that $\langle \overline{F}_{AD,\varnothing,AD} \rangle = \langle \overline{F}_{AD,\varnothing,PhF} \rangle = \langle \overline{F}_{AD,\varnothing,D} \rangle = 1$ for $p_1 = p_3 = 0$ and

$\theta = \varphi = \pi/4$, $\langle \overline{F}_{BF,\varnothing,PhF} \rangle = 1$ for $p_3 = 0$ or $p_3 = 1$ with $\theta = \varphi = \pi/4$ and $p_1 = 0$.

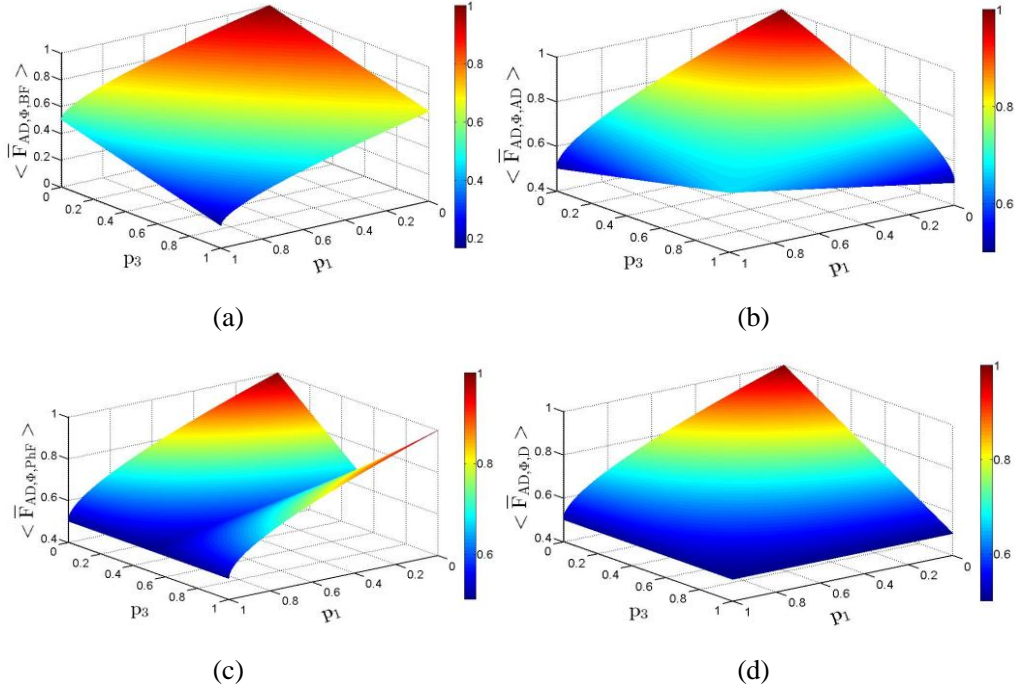

(a)                          (b)

(c)                          (d)

FIG. 5. (Color online) Optimal efficiency of the controlled RSP protocol when both Alice's qubit ($p_1$) and Charlie's qubit ($p_3$) are affected by a noisy environment. Here the qubit 1 is always subjected to the amplitude-damping (AD) noise while Charlie's qubit may suffer from several types of noise.

From Eqs. (26)-(29), we see that the efficiencies depend on the coupling terms between the two noisy rates. In the other words, such efficiencies are relative to the entanglement of environments. It is happens again for Charlie's qubit interacting with the bit-flip noise [Eq. (26)], for the optimal $\theta$ is not $\pi/4$, the less entanglement of environments leads to a better performance for the controlled RSP protocol.

Similarly, in order to implement the perfect controlled RSP by using a non-maximally three-qubit pure entangled state under the noise environments in this case, we set $\langle \overline{F}_{AD,\varnothing,BF} \rangle = 1$ in Eq. (26). Thus we have

$$p_3 = \frac{2 - \sin^2\theta + 2p_1 - 2\sin(2\theta)\sqrt{1-p_1}}{\sin^2\theta - 2 - p_1}. \tag{30}$$

The numerical results $p_3$ of Eq. (30) is shown in Fig. 6. We see that one can obtain $\langle \overline{F}_{AD,\varnothing,BF} \rangle = 1$ for $0 \leq p_1 \leq 0.15$, $0 < p_3 < 0.45$, $0.75 \leq \theta \leq 1.08$ [rad].

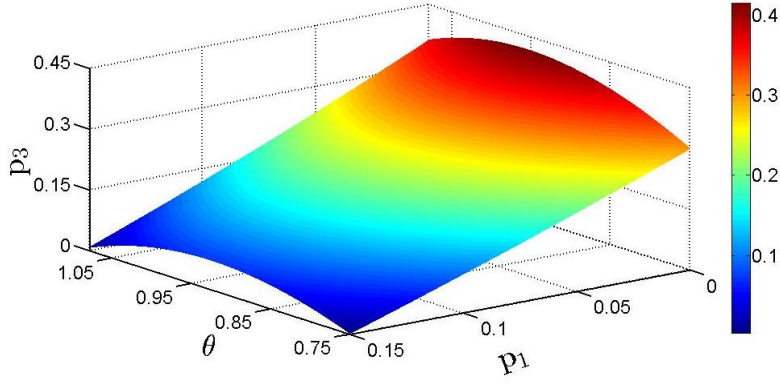

FIG. 6. (Color online) The relations of realizing the perfect controlled RSP between the noisy rates and measured angle, where the qubit 1 is subjected to the amplitude-damping (AD) noise while Charlie's qubit may suffer from bit-flip (BF) noise: $0 \leq p_1 \leq 0.15$, $0 < p_3 < 0.45$, $0.75 \leq \theta \leq 1.08$ [rad], and $\langle \overline{F}_{AD,\varnothing,BF} \rangle = 1$

When the qubit 1 is subjected to the phase-flip noise and Charlie's qubit is subjected to the four different types of noise as shown in the appendix, the optimal efficiencies are

$$\langle \overline{F}_{PhF,\varnothing,BF} \rangle = \frac{2}{3} - \frac{p_3}{3} + \frac{1}{3}|1-2p_1|\sin(2\theta)\sin(2\varphi), \tag{31}$$

$$\langle \overline{F}_{PhF,\varnothing,AD} \rangle = \frac{2}{3} - \frac{p_3}{3}\sin^2\theta + \frac{1}{3}|1-2p_1|\sqrt{1-p_3}\sin(2\theta)\sin(2\varphi), \tag{32}$$

$$\langle \overline{F}_{PhF,\varnothing,PhF} \rangle = \frac{2}{3} + \frac{1}{3}|(1-2p_1)(1-2p_3)|\sin(2\theta)\sin(2\varphi), \tag{33}$$

$$\langle \overline{F}_{PhF,\varnothing,D} \rangle = \frac{2}{3} - \frac{p_3}{6} + \frac{1}{3}|1-2p_1|(1-p_3)\sin(2\theta)\sin(2\varphi). \tag{34}$$

In Eqs. (31)-(34), the optimal efficiencies are occurred at $\theta = \varphi = \pi/4$. In Fig. 7, we plot the maximum efficiencies for different values of the noisy rate $p_1$ and $p_3$ in terms of Eqs. (31)-(34), where $0 \leq p_1 \leq 1$, $0 \leq p_3 \leq 1$, and $\theta = \varphi = \pi/4$.

For $p_1 < 0.5$, we find that $\langle \overline{F}_{PhF,\varnothing,BF} \rangle$ and $\langle \overline{F}_{PhF,\varnothing,D} \rangle$ reduce with increasing of the noisy rate $p_3$. The results imply that the average fidelities decrease with increasing of noisy rate. Differently from $\langle \overline{F}_{PhF,\varnothing,PhF} \rangle$, $\langle \overline{F}_{PhF,\varnothing,BF} \rangle$ and $\langle \overline{F}_{PhF,\varnothing,D} \rangle$ are divided into two regions, which raise for $p_3 > 0.5$ and reduce for $p_3 < 0.5$.

It is interesting that $\langle \overline{F}_{PhF,\varnothing,PhF} \rangle$ is divided into four regions, which is reduces for $p_3 < 0.5$ and $p_1 < 0.5$ and raises for $p_3 > 0.5$ and $p_1 > 0.5$. It is surprised that under the situations of $p_3 > 0.5$ and $p_1 > 0.5$ due to the entanglement of environments as shown in Eq. (33), the average fidelity $\langle \overline{F}_{PhF,\varnothing,PhF} \rangle$ becomes bigger with increasing of noisy rate $p_3$. The results show that more noise means more efficiency in the case. By putting Charlie's qubit in a noisy environments described by the phase-flip map, thus, we can raise the efficiency of the protocol beyond the classical limit with the value 2/3 [39].

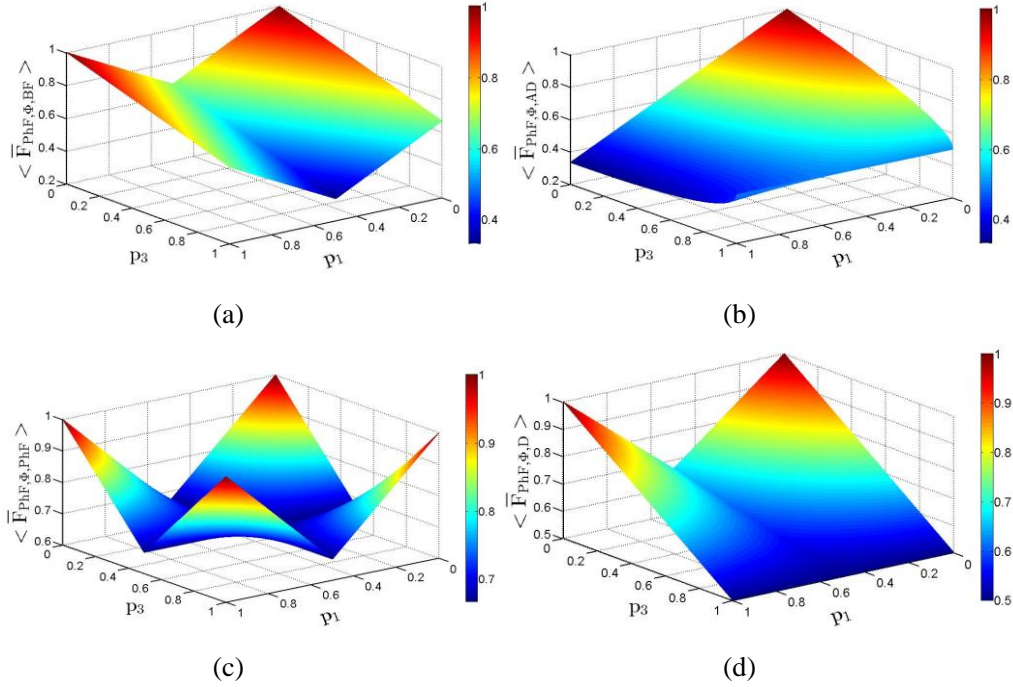

(a) (b)

(c) (d)

FIG. 7. (Color online) Optimal efficiency of the controlled RSP protocol when both Alice's qubit ($p_1$) and Charlie's qubit ($p_3$) are affected by a noisy environment. Here the qubit 1 is always subjected to the phase-flip (PhF) noise while Charlie's qubit may suffer from several types of noise.

Next, we discuss the optimal efficiencies under case of the qubit 1 with the depolarizing noise. Under this situation, the optimal efficiencies can be expressed as

$$\langle \overline{F}_{D,\varnothing,BF} \rangle = \frac{2}{3} - \frac{p_1}{6} - \frac{p_3}{3} + \frac{p_1 p_3}{3} + \frac{1}{3}(1-p_1)\sin(2\theta)\sin(2\varphi), \tag{35}$$

$$\langle \overline{F}_{D,\varnothing,AD} \rangle = \frac{2}{3} - \frac{p_1}{6} - \frac{p_3}{3}(1-p_1)\sin^2\theta + \frac{1}{3}(1-p_1)\sqrt{1-p_3}\sin(2\theta)\sin(2\varphi), \tag{36}$$

$$\langle \overline{F}_{D,\varnothing,PhF} \rangle = \frac{2}{3} - \frac{p_1}{6} + \frac{1}{3}(1-p_1)|1-2p_3|\sin(2\theta)\sin(2\varphi), \tag{37}$$

$$\langle \overline{F}_{D,\varnothing,D} \rangle = \frac{2}{3} - \frac{p_1}{6} - \frac{p_3}{6} + \frac{p_1 p_3}{6} + \frac{1}{3}(1-p_1)(1-p_3)\sin(2\theta)\sin(2\varphi). \tag{38}$$

From Eqs. (35)-(38), the optimal efficiencies are occurred at $\theta = \varphi = \pi/4$. In this case, the environments all are entangle. In Fig. 8, we plot the maximum efficiencies for different values of the noisy rate $p_1$ and $p_3$ in terms of Eqs. (35)-(38), where $0 \leq p_1 \leq 1$, $0 \leq p_3 \leq 1$, and $\theta = \varphi = \pi/4$.

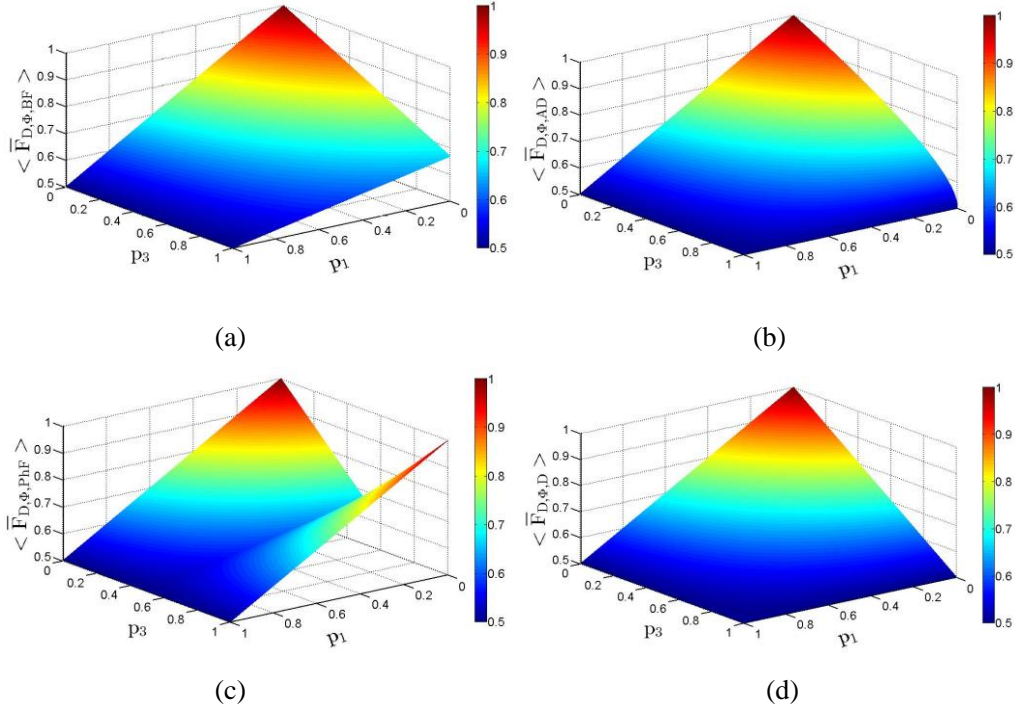

(a)      (b)

(c)      (d)

FIG. 8. (Color online) Optimal efficiency of the controlled RSP protocol when both Alice's qubit ($p_1$) and Charlie's qubit ($p_3$) are affected by a noisy environment. Here the qubit 1 is always subjected to the depolarizing (D) noise while Charlie's qubit may suffer from several types of noise.

In Fig. 8, with increasing the noisy rate $p_1$, $\langle \overline{F}_{D,\varnothing,AD} \rangle$, $\langle \overline{F}_{D,\varnothing,D} \rangle$ and $\langle \overline{F}_{D,\varnothing,BF} \rangle$ decrease beside the $\langle \overline{F}_{D,\varnothing,PhF} \rangle$, where $\langle \overline{F}_{D,\varnothing,PhF} \rangle$ is divided into two regions. In the case of $p_3 < 0.5$, $\langle \overline{F}_{D,\varnothing,PhF} \rangle$ reduces and increases for $p_3 > 0.5$. For values $p_1$ greater than $\approx 0.7$ we see that any values of average fidelities are below the classical 2/3 limit. In this situation, the RSP protocol is not possible. Therefore, it is necessary to controlling the noisy rate $p_1$ and $p_3 < 0.5$ in processing of the protocol.

**B. Noise in Bob's qubit**

Let us discuss the qubit 2 interacting with the amplitude damping noise, while Charlie's qubit can suffer any one of the four kinds of noise as shown in the appendix. The optimal efficiencies

are given by

$$\langle \overline{F}_{\varnothing,AD,BF} \rangle = \frac{2}{3} - \frac{p_2}{3} - \frac{p_3}{3} + \frac{p_2 p_3}{6} + \left(\frac{2p_2}{3} - \frac{p_2 p_3}{3}\right)\sin^2\theta + \frac{\sqrt{1-p_2}}{3}\sin(2\theta)\sin(2\varphi), \quad (39)$$

$$\langle \overline{F}_{\varnothing,AD,AD} \rangle = \frac{2}{3} - \frac{p_3}{6}[1-\cos(2\theta)] + \frac{1}{3}\sqrt{(1-p_2)(1-p_3)}\sin(2\theta)\sin(2\varphi), \quad (40)$$

$$\langle \overline{F}_{\varnothing,AD,PhF} \rangle = \frac{2}{3} - \frac{p_2}{3} + \frac{2p_2}{3}\sin^2\theta + \frac{\sqrt{1-p_2}}{3}|1-2p_3|\sin(2\theta)\sin(2\varphi), \quad (41)$$

$$\langle \overline{F}_{\varnothing,AD,D} \rangle = \frac{2}{3} - \frac{p_3}{6} - \frac{p_2}{3} + \frac{1}{12}p_2 p_3 + \left(\frac{2}{3} - \frac{p_3}{6}\right)p_2\sin^2\theta + \frac{\sqrt{1-p_2}}{3}(1-p_3)\sin(2\theta)\sin(2\varphi). \quad (42)$$

In Eqs. (39)-(42), the optimal efficiencies are occurred at $\theta = \varphi = \pi/4$, where the coupling terms of environments are emerged in all cases. In Fig. 9, we plot the maximum efficiencies for different values of the noisy rate $p_2$ and $p_3$ in terms of Eqs. (39)-(42), where $0 \leq p_2 \leq 1$, $0 \leq p_3 \leq 1$, and $\theta = \varphi = \pi/4$.

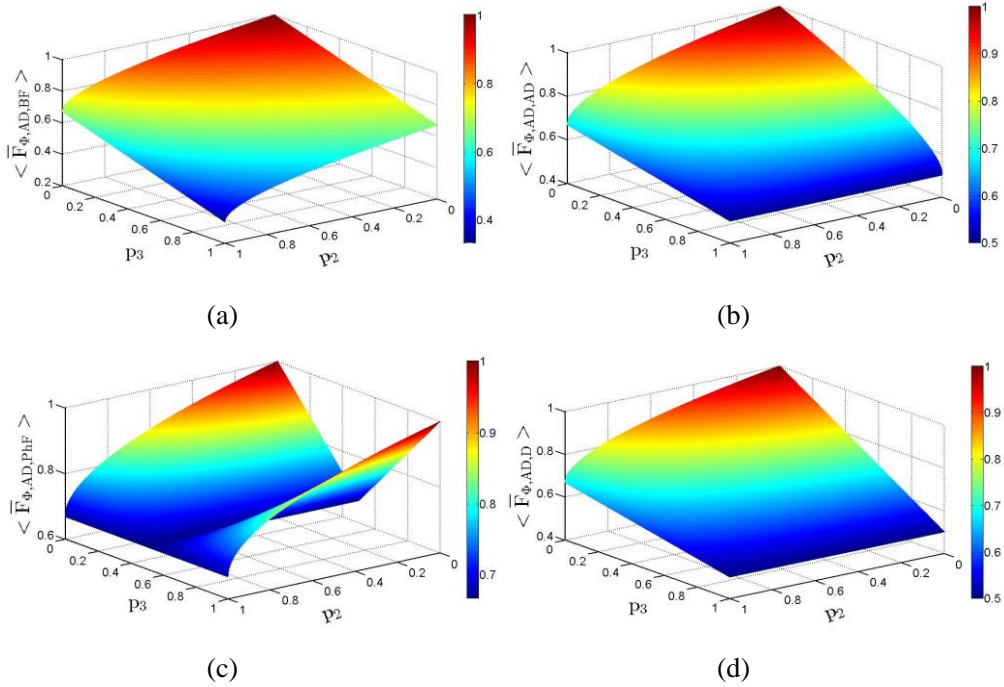

FIG. 9. (Color online) Optimal efficiency of the controlled RSP protocol when both Bob's qubit ($p_2$) and Charlie's qubit ($p_3$) are affected by a noisy environment. Here the qubit 2 is always subjected to the amplitude-damping (AD) noise while Charlie's qubit may suffer from several types of noise.

In Fig. 9, with increasing the noisy rate $p_2$, $\langle \overline{F}_{\varnothing,AD,BF} \rangle$, $\langle \overline{F}_{\varnothing,AD,AD} \rangle$ and $\langle \overline{F}_{\varnothing,AD,D} \rangle$ decrease beside the $\langle \overline{F}_{\emptyset,AD,PhF} \rangle$, where $\langle \overline{F}_{\emptyset,AD,PhF} \rangle$ is divided into two regions. In the case of $p_3 < 0.5$, $\langle \overline{F}_{\emptyset,AD,PhF} \rangle$ reduces and increases for $p_3 > 0.5$.

We find $\langle \overline{F}_{\emptyset,AD,BF} \rangle = \langle \overline{F}_{\emptyset,AD,AD} \rangle = \langle \overline{F}_{\emptyset,AD,PhF} \rangle = \langle \overline{F}_{\emptyset,AD,D} \rangle = 1$ for $p_2 = p_3 = 0$ with $\theta = \varphi = \pi/4$. In addition, $\langle \overline{F}_{\emptyset,AD,PhF} \rangle = 1$ for $p_2 = 0$, $p_3 = 1$ and $\theta = \varphi = \pi/4$.

Under the case of qubit 2 interacting with the depolarizing noise and Charlie's qubit that lies in one of the four different types of noisy environments as shown in the appendix. The optimal efficiencies in those four cases are

$$\langle \overline{F}_{\emptyset,D,BF} \rangle = \frac{2}{3} - \frac{p_3}{3} + \frac{1}{3}(1-p_2)\sin(2\theta)\sin(2\varphi), \tag{43}$$

$$\langle \overline{F}_{\emptyset,D,AD} \rangle = \frac{2}{3} - \frac{p_3}{6}[1-\cos(2\theta)] + \frac{1}{3}(1-p_2)\sqrt{1-p_3}\sin(2\theta)\sin(2\varphi), \tag{44}$$

$$\langle \overline{F}_{\emptyset,D,PhF} \rangle = \frac{2}{3} + \frac{1}{3}(1-p_2)|1-2p_3|\sin(2\theta)\sin(2\varphi), \tag{45}$$

$$\langle \overline{F}_{\emptyset,D,D} \rangle = \frac{2}{3} - \frac{p_3}{6} + \frac{1}{3}(1-p_2)(1-p_3)\sin(2\theta)\sin(2\varphi). \tag{46}$$

In Eqs. (43)-(46), the optimal efficiencies are occurred at $\theta = \varphi = \pi/4$. In Fig. 10, we plot the maximum efficiencies different values of the noisy rate $p_2$ and $p_3$ in terms of Eqs. (43)-(46), where $0 \leq p_2 \leq 1$, $0 \leq p_3 \leq 1$, and $\theta = \varphi = \pi/4$.

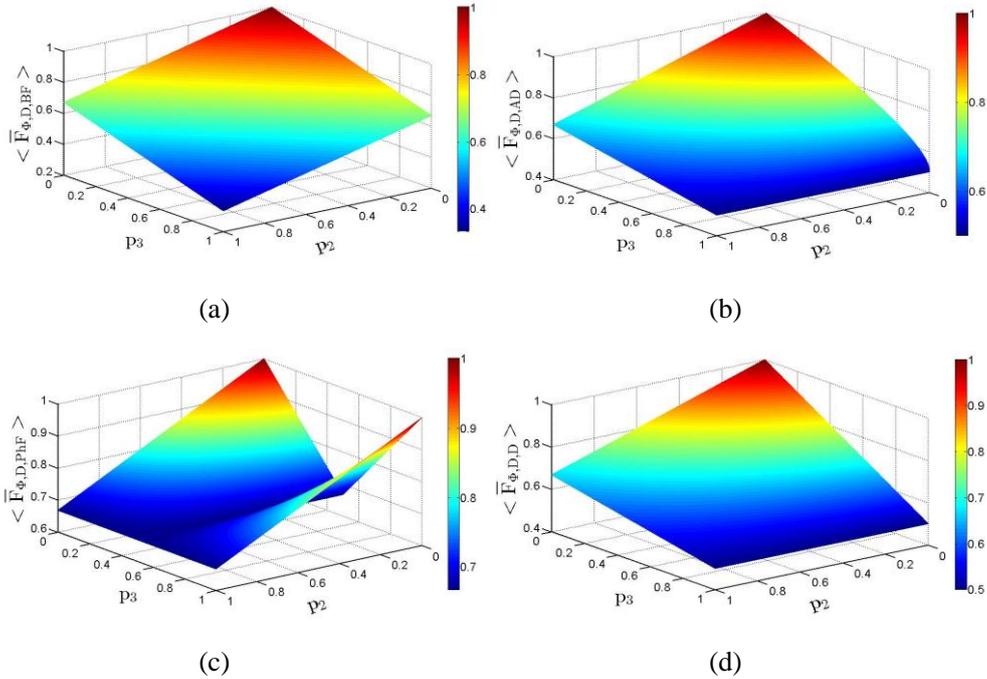

(a)　　　　　　　　　　　　　　(b)

(c)　　　　　　　　　　　　　　(d)

FIG. 10. (Color online) Optimal efficiency of the controlled RSP protocol when both Bob's qubit ($p_2$) and Charlie's qubit ($p_3$) are affected by a noisy environment. Here the qubit 2 is always subjected to the depolarizing (D) noise while Charlie's qubit may suffer from several types of noise.

In Fig. 10, with increasing the noisy rate $p_2$, $\langle \overline{F}_{\varnothing,D,BF} \rangle$, $\langle \overline{F}_{\varnothing,D,AD} \rangle$ and $\langle \overline{F}_{\varnothing,D,D} \rangle$ decrease beside the $\langle \overline{F}_{\varnothing,D,PhF} \rangle$, where $\langle \overline{F}_{\varnothing,D,PhF} \rangle$ is divided into two regions. In the case of $p_3 < 0.5$, $\langle \overline{F}_{\varnothing,D,PhF} \rangle$ reduces and increases for $p_3 > 0.5$. By putting Charlie's qubit in a noisy environments described by the phase-flip map, thus, we can raise the efficiency of the protocol beyond the classical limit with the value 2/3.

When the qubit 2 is subjected either to the bit-flip noise or to the phase-flip noise, we find that the qualitative behavior of $\langle \overline{F}_{\varnothing,BF,Y} \rangle, (Y = \varnothing, BF, PhF, D, AD)$ is similar to Fig. 8. A direct calculation shows $\langle \overline{F}_{\varnothing,PhF,Y} \rangle = \langle \overline{F}_{PhF,\varnothing,Y} \rangle$. In other words, the qualitative behavior of $\langle \overline{F}_{\varnothing,PhF,Y} \rangle$ is the same as Fig. 7.

In the cases, we do not find an approach to improve the overall efficiency by adjusting the noise rate from Bob's qubit and one could not implement the perfect controlled RSP by using a non-maximally three-qubit pure entangled state in the cases. When Bob's qubit is subjected to the bit-flip noise and amplitude-damping noise, especially, we can perform such a protocol in the higher efficiency for low values of $p_2$. For the higher values of $p_2$, on the other hand, the phase-flip channel may be a better choose to perform the protocol.

**C. Noise in Alice's and Bob's qubits**

Next, we further investigate the scenario that the qubits 1 and 2 are subjected to the same type of noises. This scenario is useful in the controlled RSP protocol since the entangled channel is employed by Charlie for the quantum communication tasks, suppose that $p_1 = p_2 = p$, but Charlie's qubit may suffer a different type of noise.

For the qubits 1 and 2 interacting with the bit-flip noise, we have the following optimal efficiencies,

$$\langle \overline{F}_{BF,BF,BF} \rangle = \frac{2}{3} - \frac{4}{3}p + p^2 - \frac{1}{3}p_3 + \frac{2}{3}pp_3 + \frac{1}{3}(1 - 2p + 2p^2)\sin(2\theta)\sin(2\varphi), \qquad (47)$$

$$\langle \overline{F}_{BF,BF,AD} \rangle = \frac{2}{3} - \frac{4}{3}p + p^2 - \frac{p_3}{6}(1 - 2p + 2p^2)[1 - \cos(2\theta)] \qquad (48)$$
$$+ \frac{1}{3}(1 - 2p + 2p^2)\sqrt{1 - p_3} \sin(2\theta)\sin(2\varphi),$$

$$\langle \overline{F}_{BF,BF,PhF} \rangle = \frac{2}{3} - \frac{4}{3}p + p^2 + \frac{1}{3}|1 - 2p_3|(1 - 2p + 2p^2)\sin(2\theta)\sin(2\varphi), \qquad (49)$$

$$\langle \overline{F}_{BF,BF,D} \rangle = \frac{2}{3} - \frac{4}{3}p + p^2 - \frac{p_3}{6} + \frac{1}{3}pp_3 + \frac{1}{3}(1 - 2p + 2p^2)(1 - p_3)\sin(2\theta)\sin(2\varphi). \qquad (50)$$

From Eqs. (47)-(50), the optimal efficiencies depend on noisy rate $p$, $p_3$, $\theta$ and $\varphi$. And the optimal efficiencies are occurred at $\theta = \varphi = \pi/4$. In Fig. 11, we plot the maximum efficiencies different values of the noisy rate $p$ and $p_3$ in terms of Eqs. (47)-(50), where $0 \le p \le 1$, $0 \le p_3 \le 1$, and $\theta = \varphi = \pi/4$.

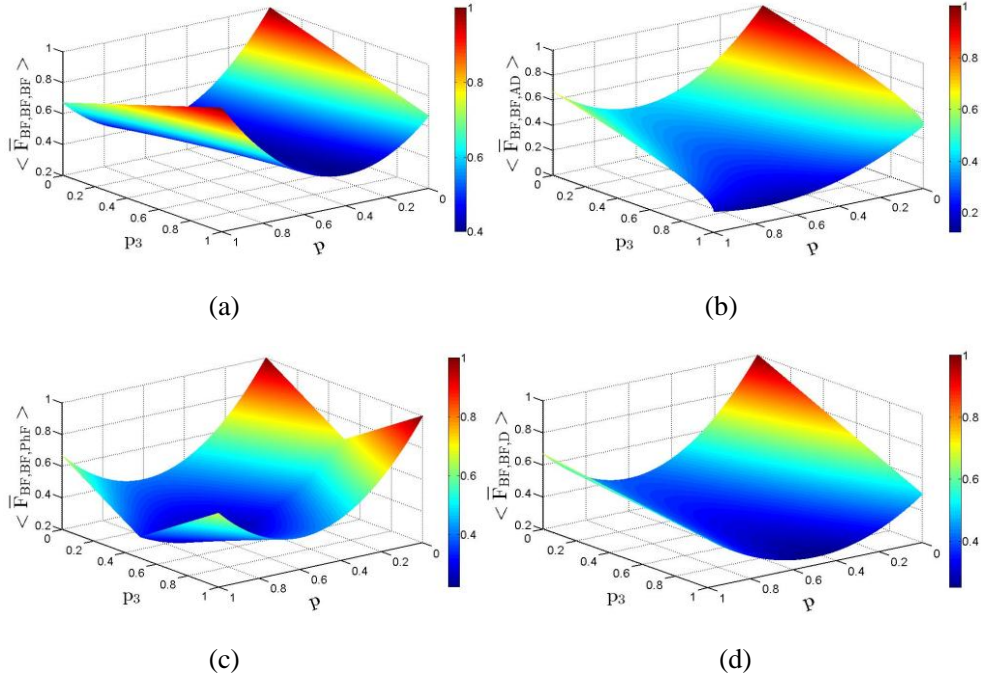

(a) (b)

(c) (d)

FIG. 11. (Color online) Optimal efficiency of the controlled RSP protocol when the qubits 1 and 2 ($p$) and Charlie's qubit ($p_3$) are affected by a noisy environment. Here the qubits 1 and 2 is always subjected to the bit-flip (BF) noise while Charlie's qubit may suffer from several types of noise.

Fig. 11 shows the dynamic behaviors of Eqs. (43)-(46) for the different noisy rate $p$ and $p_3$. In the region of $p > 0.9$, we see that more noise means more efficiency. By adding more noise to Charlie's qubit in a noisy environment of the bit-flip map, we can increase the efficiency of protocol, where the following optimal efficiency increases with increasing the noisy rate $p_3 > 0.5$ above the classical limit.

Under the case of qubits 1 and 2 interacting with the amplitude-damping noise, the optimal efficiencies are

$$\langle \overline{F}_{AD,AD,BF} \rangle = \frac{1}{6}(2-p_3)(2-2p+p^2) + \frac{1}{3}p^2(1+p_3)\sin^2\theta + \frac{1}{3}(1-p)\sin(2\theta)\sin(2\varphi), \tag{51}$$

$$\langle \overline{F}_{AD,AD,AD} \rangle = \frac{2}{3} - \frac{1}{3}(4p - 3p^2 + p_3 - 2pp_3)\sin^2\theta + \frac{1}{3}(1-p)\sqrt{1-p_3}\sin(2\theta)\sin(2\varphi), \tag{52}$$

$$\langle \overline{F}_{AD,AD,PhF} \rangle = \frac{2}{3} - \frac{2}{3}p + \frac{1}{3}p^2 + \frac{1}{3}p^2\sin^2\theta + \frac{1}{3}(1-p)|1-2p_3|\sin(2\theta)\sin(2\varphi), \tag{53}$$

$$\langle \overline{F}_{AD,AD,D} \rangle = \frac{1}{12}(4-p_3)(2-2p+p^2) + \frac{1}{6}p^2(2+p_3)\sin^2\theta + \frac{1}{3}(1-p)(1-p_3)\sin(2\theta)\sin(2\varphi). \tag{54}$$

From Eqs. (51)-(54), we see the optimal efficiencies depend on noisy rate $p$, $p_3$, $\theta$ and $\varphi$. And the optimal efficiencies are occurred at $\theta = \varphi = \pi/4$. In Fig. 12, we plot the optimal efficiencies different values of the noisy rate $p$ and $p_3$ in terms of Eqs. (51)-(54), where $0 \leq p \leq 1$, $0 \leq p_3 \leq 1$, and $\theta = \varphi = \pi/4$.

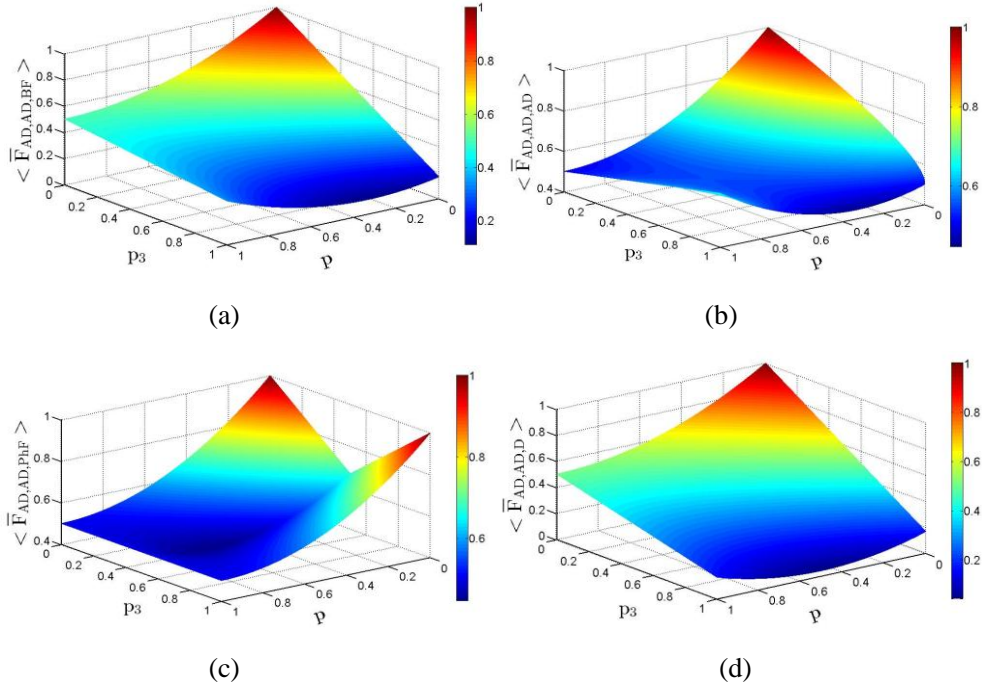

(a)      (b)

(c)      (d)

FIG. 12. (Color online) Optimal efficiency of the controlled RSP protocol when the qubits 1 and 2 ($p$) and Charlie's qubit ($p_3$) are affected by a noisy environment. Here the qubits 1 and 2 is always subjected to amplitude-damping (AD) noise while Charlie's qubit may suffer from several types of noise.

In Fig. 12, with increasing the noisy rate $p$, $\langle \overline{F}_{AD,AD,BF} \rangle$, $\langle \overline{F}_{AD,AD,AD} \rangle$ and $\langle \overline{F}_{AD,AD,D} \rangle$ decrease beside the $\langle \overline{F}_{AD,AD,PhF} \rangle$, where $\langle \overline{F}_{AD,AD,PhF} \rangle$ is divided into two regions. In the case of $p_3 < 0.5$, $\langle \overline{F}_{AD,AD,PhF} \rangle$ reduces and increases for $p_3 > 0.5$.

When the qubits 1 and 2 are subjected to the phase-flip noise and Charlie's qubit suffer any one of the four kinds of noise given in Sec. 3. The optimal efficiencies become

$$\langle \overline{F}_{PhF,PhF,BF} \rangle = \frac{1}{3}(2-p_3)(1-2p+2p^2) + \frac{1}{3}(1-2p+2p^2)\sin(2\theta)\sin(2\varphi), \tag{55}$$

$$\langle \overline{F}_{PhF,PhF,AD} \rangle = \frac{2}{3}(1-2p+2p^2) - \frac{p_3}{6}[1-\cos(2\theta)] + \frac{1}{3}(1-2p+2p^2)\sqrt{1-p_3}\sin(2\theta)\sin(2\varphi), \tag{56}$$

$$\langle \overline{F}_{PhF,PhF,PhF} \rangle = \frac{2}{3}(1-2p+2p^2) + \frac{1}{3}(1-2p+2p^2)|1-2P_3|\sin(2\theta)\sin(2\varphi), \tag{57}$$

$$\langle \overline{F}_{PhF,PhF,D} \rangle = \frac{1}{6}(4-p_3)(1-2p+2p^2) + \frac{1}{3}(1-p_3)(1-2p+2p^2)\sin(2\theta)\sin(2\varphi). \tag{58}$$

From Eqs. (55)-(58), the optimal efficiencies are occurred at $\theta = \varphi = \pi/4$. In Fig. 13, we plot the maximum efficiencies different values of the noisy rate $p$ and $p_3$ in terms of Eqs. (55)-(58), where $0 \le p \le 1$, $0 \le p_3 \le 1$, and $\theta = \varphi = \pi/4$.

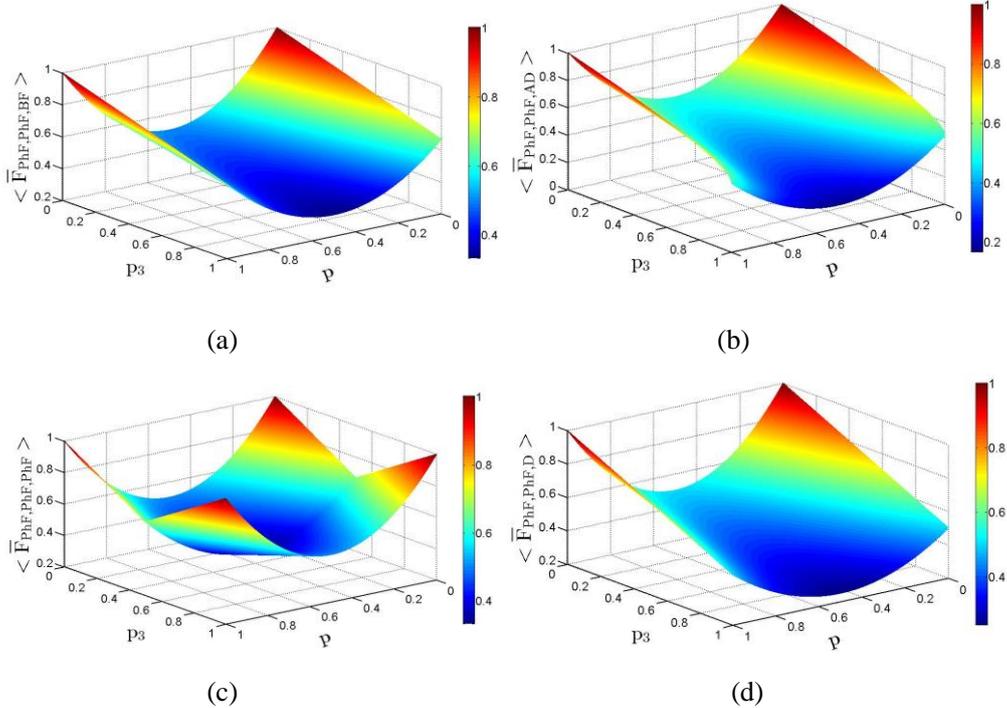

(a)　　　　　　　　　　　　　(b)

(c)　　　　　　　　　　　　　(d)

FIG. 13. (Color online) Optimal efficiency of the controlled RSP protocol when the qubits 1 and 2 ($p$) and Charlie's qubit ($p_3$) are affected by a noisy environment. Here the qubits 1 and 2 is always subjected to the phase-flip (PhF) noise while Charlie's qubit may suffer from several types of noise.

Fig. 13 shows the dynamic behaviors for the different noisy rate $p$ and $p_3$. Here $\langle \overline{F}_{PhF,PhF,BF} \rangle$ is divided into four regions, which is reduces for $p_3 < 0.5$ and $p < 0.5$ and raises for $p_3 > 0.5$ and $p > 0.5$. It is surprised that under the situations of $p_3 > 0.5$ and $p > 0.5$, the average fidelity $\langle \overline{F}_{PhF,PhF,BF} \rangle$ becomes bigger with increasing of noisy rate $p_3$. The results show that more noise means more efficiency in the case. By putting Charlie's qubit in a noisy environments described by the phase-flip map, thus, we can raise the efficiency of the protocol beyond the classical limit with the value 2/3.

Assuming that the qubits 1 and 2 are subjected to the depolarizing noise and Charlie's qubit is subjected to the four different types of noise, the optimal efficiencies are

$$\langle \overline{F}_{D,D,BF} \rangle = \frac{2}{3} - p + \frac{11}{24}p^2 + \frac{1}{2}pp_3 - \frac{1}{3}p_3 - \frac{1}{6}p^2 p_3 + \left(\frac{1}{3} - \frac{1}{2}p + \frac{1}{4}p^2\right)\sin(2\theta)\sin(2\varphi), \tag{59}$$

$$\begin{aligned}\langle \overline{F}_{D,D,AD} \rangle &= \frac{2}{3} - p + \frac{11}{24}p^2 - \frac{1}{6}(2p_3 - 3pp_3 + p^2 p_3)[1 - \cos(2\theta)] \\ &+ \frac{1}{3}\left(1 - \frac{3}{2}p + \frac{3}{4}p^2\right)\sqrt{1-p_3}\sin(2\theta)\sin(2\varphi),\end{aligned} \tag{60}$$

$$\langle \overline{F}_{D,D,PhF} \rangle = \frac{2}{3} - p + \frac{11}{24}p^2 + \left(\frac{1}{3} - \frac{1}{2}p + \frac{1}{4}p^2\right)|1-2p_3|\sin(2\theta)\sin(2\varphi), \tag{61}$$

$$\begin{aligned}\langle \overline{F}_{D,D,D} \rangle &= \frac{2}{3} - p + \frac{11}{24}p^2 + \frac{1}{4}pp_3 - \frac{1}{6}p_3 - \frac{1}{12}p^2 p_3 \\ &+ \left(\frac{1}{3} - \frac{1}{2}p + \frac{1}{4}p^2\right)(1-p_3)\sin(2\theta)\sin(2\varphi).\end{aligned} \tag{62}$$

From Eqs. (59)-(62), we see the optimal efficiencies depend on noisy rate $p$, $p_3$, $\theta$ and $\varphi$. And the optimal efficiencies are occurred at $\theta = \varphi = \pi/4$. In Fig. 14, we plot the optimal efficiencies different values of the noisy rate $p$ and $p_3$ in terms of Eqs. (59)-(62), where $0 \le p \le 1$, $0 \le p_3 \le 1$, and $\theta = \varphi = \pi/4$.

The results show that by putting Charlie's qubit in a noisy environment given in Sec. 3, thus, we cannot raise the efficiency of the protocol.

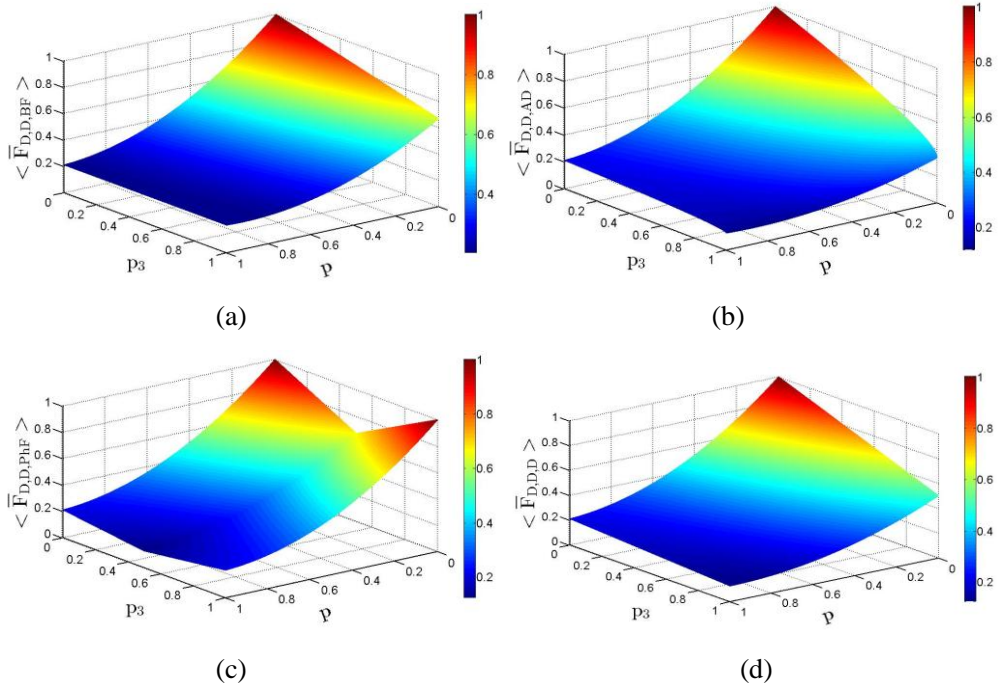

FIG. 14. (Color online) Optimal efficiency of the controlled RSP protocol when the qubits 1 and 2 ($p$) and Charlie's qubit ($p_3$) are affected by a noisy environment. Here the qubits 1 and 2 is always subjected to the depolarizing (D) noise while Charlie's qubit may suffer from several types of noise.

## 5. Conclusions

In summary, we investigate how the efficiency of the controlled RSP protocol is affected by all possibly noisy environments, which includes all decay ways, i.e., the bit-flip noise, amplitude-damping noise, phase-flip noise, and depolarizing noise channels, in the realistic quantum communication protocols. We also studied the different scenarios with one, two, or all three qubits in the noisy environment for the protocol. We find an approach to keep the perfect controlled RSP in the presence of noise by controlling the entanglement of environment and measured angle.

The results show that when the noise is present in both the desired qubit and the quantum channel, the optimal efficiency is related to between the environmental entanglement and qubits. We find that the less such an entanglement is, the more the efficiency of average fidelity is. For the quantum channels interacting with the amplitude-damping noise, we show that by controlling the noisy rate closed to one, an approximately perfect efficiency of the controlled RSP protocol can be obtained, and one could realize the perfect controlled RSP in noisy environments.

We further showed that such an efficiency depend on the noisy rate and the initial state. When the qubit 1 lies in the bit-flip noise and Charlie's qubit at the same time, the efficiency is considerably greater in comparison with the situation of only the qubit 1 in this type of noise. This kind of behavior was also observed in Ref. [35] when the channel qubits are subjected to the amplitude-damping noise.

By choosing the kind of noise interacting with the qubits, we find that the optimal combination in the noisy environments can lead to the greatest efficiencies. In many situations, Alice, Bob and Charlie should subject their qubits to the kind of noise in order to get the better scheme to perform the perfect controlled RSP protocol. A potentially feasible approach to the optimal efficiency is obtained by putting one of the qubits be sent to different kinds of noise for a longer time than the other one.

**Acknowledgements** This work was supported in part by the National Basic Research Program 973 of China under Grant No.2011CB808101, the National Natural Science Foundation of China under Grant Nos. 61125503, 61235009, 61505189, 61205110, and 11564018, the Foundation for Development of Science and Technology of Shanghai under Grant No. 13JC1408300, Presidential Foundation of the China Academy of Engineering Physics (Grant No.201501023), the Natural Science Foundation of Jiangxi Province, China (Grant No. 20142BAB202005), the Research Foundation of the Education Department of Jiangxi Province (No. GJJ150339), and in part by the Innovative Foundation of Laser Fusion Research Center.

# Appendix

**A. Bit-flip noise**

The bit-flip noise changes a qubit state from $|0\rangle$ to $|1\rangle$ or from $|1\rangle$ to $|0\rangle$ with a probability $p$ and is frequently used in the theory of quantum-error correction. The associated Kraus operators are given by

$$E_1 = \begin{pmatrix} \sqrt{1-p} & 0 \\ 0 & \sqrt{1-p} \end{pmatrix}, E_2 = \begin{pmatrix} 0 & \sqrt{p} \\ \sqrt{p} & 0 \end{pmatrix}. \tag{63}$$

**B. Amplitude-damping noise**

The amplitude-damping noise channel allows us to describe the decay of a two-level system due to spontaneous emission of a phonon. This process is accompanied with the loss of energy and can be described by the Kraus operators,

$$E_1 = \begin{pmatrix} 1 & 0 \\ 0 & \sqrt{1-p} \end{pmatrix}, E_2 = \begin{pmatrix} 0 & \sqrt{p} \\ 0 & 0 \end{pmatrix}. \tag{64}$$

The quantity $p$ is regarded as a decay probability from the excited to the ground state for a two-level system.

**C. Phase-flip noise**

The phase-flip noise channel has no classical analog because it describes the loss of quantum information without loss of energy. The quantum information corresponding to the ability of a system to produce quantum interferences hence is described by the off-diagonal elements of a density matrix. Phase-flip map can be occurred in the phase kicks or scattering processes. Such a channel can be modeled by the following Kraus operators,

$$E_1 = \begin{pmatrix} \sqrt{1-p} & 0 \\ 0 & \sqrt{1-p} \end{pmatrix}, E_2 = \begin{pmatrix} \sqrt{p} & 0 \\ 0 & -\sqrt{p} \end{pmatrix}. \tag{65}$$

**D. Depolarizing noise**

The depolarizing noise channel is a decoherent model. The Kraus operators including all possible decay ways for the depolarizing channel are given by

$$E_1 = \begin{pmatrix} \sqrt{1-3p/4} & 0 \\ 0 & \sqrt{1-3p/4} \end{pmatrix}, E_2 = \begin{pmatrix} 0 & \sqrt{p/4} \\ \sqrt{p/4} & 0 \end{pmatrix}\sigma_x, E_3 = \begin{pmatrix} 0 & -i\sqrt{p/4} \\ i\sqrt{p/4} & 0 \end{pmatrix},$$
$$E_4 = \begin{pmatrix} \sqrt{p/4} & 0 \\ 0 & -\sqrt{p/4} \end{pmatrix}. \tag{66}$$